\documentclass[useAMS,usenatbib]{mn2e}
\usepackage{txfonts}
\usepackage{natbib}
\usepackage[dvips]{graphicx}

\usepackage{hyperref}
\usepackage[usenames,dvipsnames]{color}
\definecolor{menublue}{rgb}{0.0,0.0,0.5}
\definecolor{citegreen}{rgb}{0.0,1.0,0.0}
\definecolor{urlred}{rgb}{1.0,0.0,0.0}
\hypersetup{bookmarksopen,pdfstartview={FitH},colorlinks=true, breaklinks=true,menucolor=menublue,urlcolor=urlred,citecolor=citegreen,linkcolor=blue}
\usepackage[all]{hypcap}

\def\del#1{{}}

\onecolumn

\newcommand{\ltsima}{$\; \buildrel < \over \sim \;$}
\newcommand{\lsim}{\lower.5ex\hbox{\ltsima}}
\newcommand{\gtsima}{$\; \buildrel > \over \sim \;$}
\newcommand{\gsim}{\lower.5ex\hbox{\gtsima}}
\newcommand{\bra}{\langle}
\newcommand{\ket}{\rangle}

\newcommand{\dd}{\mathrm{d}}

\newcommand{\vecx}{\bmath{x}}

\newcommand{\vecl}{\bmath{\ell}}

\newcommand{\trace}{\mathrm{tr}}

\newcommand{\chip}{{\chi^\prime}}

\newcommand{\dirac}{\delta_D}

\newcommand{\fnl}{f_\mathrm{NL}}
\newcommand{\gnl}{g_\mathrm{NL}}
\newcommand{\tnl}{\tau_\mathrm{NL}}
\newcommand{\veck}{\bmath{k}}

\title[weak convergence extreme values]
{Extreme value statistics of the weak lensing convergence:\\ 1. primordial non-Gaussianities}
\author[F. Capranico, A.F. Kalovidouris and B.M. Sch{\"a}fer]{
Federica Capranico\thanks{e-mail: capranico@ari.uni-heidelberg.de}, Angelos Fotios Kalovidouris and Bj{\"o}rn Malte Sch{\"a}fer\\
Astronomisches Recheninstitut, Zentrum f{\"u}r Astronomie der Universit{\"a}t Heidelberg, Philosophenweg 12, 69120 Heidelberg, Germany}

\begin{document}
\pagerange{\pageref{firstpage}--\pageref{lastpage}}
\pubyear{2013}
\maketitle
\label{firstpage}

\begin{abstract}
The subject of this paper is the investigation of inflationary non-Gaussianities of the local type with extreme value statistics of the weak lensing convergence $\kappa$. Specifically, we describe the influence of inflationary non-Gaussianities parameterised by $\fnl$ and $\gnl$ on the probability distribution $p(\kappa)\dd\kappa$ of the smoothed convergence field with a Gram-Charlier series, for which we compute the cumulants $\kappa_n$ of the smoothed convergence field as a configuration space average of the weak convergence polyspectra. We derive analytical expressions for the extreme value distribution and show that they correspond very well to direct samples of extreme values from the Gram-Charlier distribution. We show how the standard Gumbel distribution for the extreme values is recovered in the limit of large sample size. We investigate the shape and position of the extreme value distribution for $\fnl$- and $\gnl$-type non-Gaussianity and quantify the dependence on the number of available samples, leading to the inference of non-Gaussianity parameters from observed extreme values. From the observation of single extreme values in the EUCLID weak lensing survey is is possible to place constraints on $\fnl$ and $\gnl$ of the order $10^2$ and $10^5$, respectively, while $\tnl$ can not be constrained in a meaningful way.
\end{abstract}


\section{Introduction}
Cosmic inflation is a mechanism by which the early Universe underwent a period of exponential accelerated expansion and has been invoked in order to solve the flatness and horizon problems \citep{Guth:1981qf}. In addition, it provides a natural explanation for the seed fluctuations from which the cosmic large-scale structure grew by gravitational instability \citep[for reviews, see][]{Bartolo:2004uq, Wang:2013kx, Martin:2013vn, Lesgourgues:2013ys}. A very important signature of inflationary theories are the statistical properties of the perturbations they cause in the cosmic distribution of matter \citep{Bardeen:1983ly, Starobinsky:1982ve}. These fluctuations are expected to be almost Gaussian, with small deviations from Gaussianity due to violated slow-roll conditions. The most general observable of a certain inflationary model is the sequence of polyspectra which describe the fluctuations in the density field (or in the gravitational potential) in Fourier-space. Their amplitudes are given by the non-Gaussianity parameters, and we focus in this work on the lowest order parameters: $\fnl$ which characterises the bispectrum and $\gnl$ which determine the magnitude of the inflationary trispectrum. In observations of the cosmic microwave background or of the cosmic large-scale structure one aims at constraining the non-Gaussianity parameters as well as at measuring the variation of the polyspectra in their dependence on the wave vector configuration. In this way it is possible to distinguish different inflationary scenarios. 

Currently, the tightest constraints on the lowest order non-Gaussianity parameters in a non-Gaussianity model of the local type come from the analysis of the cosmic microwave background by the PLANCK surveyor, who report $\fnl = 2.7 \pm 5.8$ for the bispectrum amplitude \citep{Planck-Collaboration:2013ve, Planck-Collaboration:2013bh}. Previous studies with WMAP have found bounds on these parameters to be $-7.4\times10^5<\gnl<8.2\times10^5$ and $-0.6\times10^4<\tnl<3.3\times10^4$ \citep{Smidt:2010dq} and $-5.6\times10^5<\gnl<6.4\times10^5$ \citep{Vielva:2010cr}. Data from the large-scale structure put bounds on the non-Gaussianity parameters at similar orders of magnitude: \citet{Desjacques:2010nx} quote the range $-2.5\times10^5<\gnl<8.2\times10^5$.

In this paper we focus on constraining the non-Gaussianity parameters $\fnl$ and $\gnl$ in a local model with extreme value statistics, i.e. where the measurement consists in determining the largest (or smallest) weak lensing shear in apertures of varying size. Because $\fnl$ describes the skewness of the distribution of the weak lensing convergences and $\gnl$ the kurtosis, one would expect that those parameters influence the occurrence of extreme values of the weak lensing convergence. In contrast to the direct estimation of polyspectra our measurement averages over the configuration dependence of the non-Gaussianity model and is primarily targeted at measuring the non-Gaussianity parameters themselves rather than at distinguishing configuration dependences. The specific observable we consider is the weak lensing convergence which has the advantage of being proportional to the density field. All statistical properties of the observable, including polyspectra, will be proportional to those of the field to be investigated. We use the characteristics of the EUCLID weak lensing survey, which will reach out to redshifts of unity and cover half of the sky.

Extreme value statistics \citep[for the mathematical foundation, please refer to][]{Gumbel:1954kx, Beirlant:2004qy, Gumbel:2004vn} has been applied to a range of problems in cosmology, most notably in the "pink-elephant"-argument of massive high-redshift clusters that should not have formed in $\Lambda$CDM cosmologies at the redshifts they have been observed, and to extreme features in the cosmic microwave background such as the cold spot \citep{Cruz:2005vn, Cruz:2007zr, Vielva:2010ys}. The common motivation is a reliable description of rare events: Of course with a sufficient high number of trials one would be able to observe even very unlikely events in a Gaussian random process, but it is necessary to draw conclusions on the fundamental random process from the observation of single, unlikely events \citep{Coles:2002fk, Colombi:2011ly}. Extreme value statistics aims to provide such a description and differs from the measurement of e.g. moments of the random process in the important respect that it focuses on the asyptotic behaviour of the random process at large amplitudes instead of the core of the distribution.

In this spirit, clusters of galaxies reflecting extreme values of the underlying density field have been investigated in their power to probe the cosmological model \citep{Enqvist:2011oq, Hotchkiss:2011uq, Waizmann:2011ys, Waizmann:2012kx, Waizmann:2012vn, Davis:2011zr}, where the samples are mostly resulting from $X$-ray surveys. With these samples, statistical tests of $\Lambda$CDM or of non-Gaussian initial conditions have been carried out \citep{Cayon:2011dq, Holz:2012cr, Baldi:2011nx, Chongchitnan:2012oq, Mortonson:2011kl}. Apart from the primary application in cluster catalogues, extreme value statistics has been used in statistical analysis of the temperature pattern of the cosmic microwave background \citep{Coles:1988ve, Martinez-Gonzalez:1989bs, Larson:2005hc, Hou:2009ij, Mikelsons:2009dz} and finally to the strong lensing signal of galaxy clusters \citep{Waizmann:2012qf, Redlich:2012bh, Zitrin:2009tg}

The motivation of this paper is the question if it was possible to derive constraints on inflationary non-Gaussianities from a very simple lensing experiment: If one averages the lensing signal in patches of size angular size $\theta$ and if one derives the distribution of averaged weak lensing convergences, there will be a patch with the smallest lensing convergence and one with the largest convergence. If the underlying statistics of the convergence field exhibits non-Gaussianities from inflation, the occurrence of these extreme values of the lensing convergence will be different from those expected for a Gaussian random field. In this way, we aim to constrain non-Gaussianities not from the central part of the distribution by estimating moments but rather from the wings of the distribution by quantifying the occurence of extreme values. Because the proposed measurement is a one-point statistic, it suffers from averaging over all bi- and trispectrum configurations where sensitivity is lost, but we would like to investigate if the focus on the asymptotic behaviour of the distribution far away from the mean makes up for this loss. As the non-Gaussianity model we assume the most basic local non-Gaussianity shape, but it can in principle extended to other types of inflationary non-Gaussianity or structure formation non-Gaussianity.

After summarising the necessary cosmology background including the local model for non-Gaussianities in Sect.~\ref{sect_cosmology}, we introduce the distribution of weak lensing convergence by means of a Gram-Charlier distribution in Sect.~\ref{sect_extreme} and investigate the distribution of extreme values and quantify their sensitivity on the non-Gaussianity parameters. We summarise and discuss our results in Sect.~\ref{sect_summary}.

We present all computations for a spatially flat $w$CDM cosmology, with the specific parameter choices motivated by the recent PLANCK-results \citep{Planck-Collaboration:2013ly}: $\Omega_m = 0.3$, $n_s = 1$, $\sigma_8 = 0.8$, $\Omega_b=0.04$ and $H_0=100\: h\:\mathrm{km}/\mathrm{s}/\mathrm{Mpc}$, with $h=0.7$. The dark energy equation of state was set to be $w=-0.95$. The non-Gaussianities due to inflation are taken to be of local type and described by the two non-Gaussianity parameters $\fnl$ and $\gnl$. We derive extreme value distributions for the case of the EUCLID weak lensing survey with a median redshift of 0.9 and a solid angle of $\Delta\Omega=2\pi$ \citep{2007MNRAS.381.1018A, 2009ExA....23...17R}.

\section{cosmology}\label{sect_cosmology}

\subsection{Dark energy cosmologies}
In Friedmann-Lema{\^i}tre cosmologies with zero curvature and the matter density parameter $\Omega_m$, the Hubble function $H(a)$ is given by
\begin{equation}
\frac{H^2(a)}{H_0^2} = \frac{\Omega_m}{a^{3}} + (1-\Omega_m)\exp\left(3\int_a^1\dd\ln a\:(1+w(a))\right),
\end{equation}
where $w(a)$ is the dark energy equation of state describing the ratio between pressure and density of the dark energy fluid. Comoving distances $\chi$ can be computed from the scale factor $a$ by integration,
\begin{equation}
\chi = \int_a^1\dd a\:\frac{c}{a^2 H(a)},
\end{equation}
where the Hubble distance $\chi_H=c/H_0$ can be identified as the natural cosmological distance scale.

\subsection{CDM power spectrum}
The linear CDM density power spectrum $P(k)$ describes Gaussian fluctuations of the CDM-density field $\delta$ in Fourier space, $\bra\delta(\veck_1)\delta(\veck_2)\ket=(2\pi)^3\dirac(\veck_1+\veck_2)P(k_1)$ and this variance is diagonal if the fluctuation properties are homogeneous. Inflationary models suggest
\begin{equation}
P(k)\propto k^{n_s}T^2(k),
\end{equation}
with the transfer function $T(k)$ and the spectral index $n_s$ close to unity. $T(k)$ describes the passage of modes through horizon re-entry and is approximately given by \citet{1986ApJ...304...15B},
\begin{equation}
T(q) = \frac{\ln(1+2.34q)}{2.34q}\left(1+3.89q+(16.1q)^2+(5.46q)^3+(6.71q)^4\right)^{-\frac{1}{4}},
\label{eqn_cdm_transfer}
\end{equation}
if the matter density is low. In eqn.~(\ref{eqn_cdm_transfer}), the wave vector $q=k/\Gamma$ is substituted in units of the shape parameter $\Gamma=\Omega_mh$. A nonzero baryon density causes a small correction to $\Gamma$ \citep{1995ApJS..100..281S},
\begin{equation}
\Gamma=\Omega_m h\exp\left(-\Omega_b\left(1+\frac{\sqrt{2h}}{\Omega_m}\right)\right).
\end{equation}

The normalisation of the spectrum $P(k)$ is taken to be the variance $\sigma_8^2$ on the scale $R=8~\mathrm{Mpc}/h$,
\begin{equation}
\sigma^2_R 
= \int\frac{\dd k}{2\pi^2}\:k^2 P(k) W^2(kR)
\end{equation}
with a Fourier transformed spherical top hat filter function, $W(x)=3j_1(x)/x$. $j_\ell(x)$ is the spherical Bessel function of the first kind of order $\ell$  \citep{1972hmf..book.....A}. 

Because the focus of this paper is on large-scale, inflationary non-Gaussianities, the time-evolution of all polyspectra can be predicted from linear structure formation, where $\delta(\bmath{x},a)=D_+(a)\delta(\bmath{x},a=1)$. The linear growth function $D_+(a)$ is the growing-mode solution to the growth equation \citep{1997PhRvD..56.4439T, 1998ApJ...508..483W, 2003MNRAS.346..573L},
\begin{equation}
\frac{\dd^2D_+(a)}{\dd a^2} + \frac{1}{a}\left(3+\frac{\dd\ln H}{\dd\ln a}\right)\frac{\dd D_+(a)}{\dd a} = 
\frac{3}{2a^2}\Omega_m(a) D_+(a).
\label{eqn_growth}
\end{equation}
which is applicable as long as non-linearities in the structure formation equations are weak. From the spectrum of the CDM density fluctuations one can construction the spectrum $P_\Phi(k)$ of the gravitational potential, $\bra\Phi(\veck_1)\Phi(\veck_2)\ket = (2\pi)^3\dirac(\veck_1+\veck_2)\:P_\Phi(k_1)$,
\begin{equation}
P_\Phi(k) = \left(\frac{3\Omega_m}{2\chi_H^2}\right)^2\:k^{n_s-4}\:T(k)^2
\end{equation}
by application of the comoving Poisson equation $\Delta\Phi = 3\Omega_m/(2\chi_H^2)\delta$. We focus on large angular scales, where most of the lensing signal is generated by linear structures, and extend the CDM-spectrum to nonlinear scales in some cases, by employing a nonlinear transfer function derived by \citet{2003MNRAS.341.1311S}.

\subsection{Primordial non-Gaussianities}
Non-Gaussianities of the local type are introduced as quadratic and cubic perturbations of the potential at a given point $\vecx$ \citep{Gangui:1994kx, Verde:2000uq, Komatsu:2001fk},
\begin{equation}
\Phi(\vecx) \rightarrow \Phi(\vecx) + \fnl\left(\Phi^2(\vecx) - \bra\Phi^2\ket\right) + \gnl\left(\Phi^3(\vecx)-3\bra\Phi^2\ket\Phi(\vecx)\right),
\end{equation}
with two parameters $\fnl$ and $\gnl$, which lead in Fourier-space to a bispectrum $\bra\Phi(\veck_1)\Phi(\veck_2)\Phi(\veck_3)\ket = (2\pi)^3\dirac(\veck_1+\veck_2+\veck_3)\:B_\Phi(\veck_1,\veck_2,\veck_3)$,
\begin{equation}
B_\Phi(\veck_1,\veck_2,\veck_3) = 2\fnl\:\left(\frac{3\Omega_m}{2\chi_H^2}\right)^3\:
\left((k_1k_2)^{n_s-4} + (k_2k_3)^{n_s-4} + (k_1k_3)^{n_s-4}\right)\:
T(k_1)T(k_2)T(k_3),
\end{equation}
and a corresponding trispectrum $\bra\Phi(\veck_1)\Phi(\veck_2)\Phi(\veck_3)\Phi(\veck_4)\ket = (2\pi)^3\dirac(\veck_1+\veck_2+\veck_3+\veck_4)\:T_\Phi(\veck_1,\veck_2,\veck_3,\veck_4)$,
\begin{equation}
T_\Phi(\veck_1,\veck_2,\veck_3,\veck_4) = 6\gnl\:\left(\frac{3\Omega_m}{2\chi_H^2}\right)^4\:
\left((k_1k_2k_3)^{n_s-4} + (k_1k_2k_4)^{n_s-4} + (k_1k_3k_4)^{n_s-4} + (k_2k_3k_4)^{n_s-4}\right)\:
T(k_1)T(k_2)T(k_3)T(k_4).
\end{equation}
The normalisation of each mode $\Phi(\veck)$ is set to be consistent with the normalisation $\sigma_8$ of the CDM-spectrum $P(k)$.

\subsection{Weak gravitational lensing}
Weak gravitational lensing refers to the shape distortions of light bundles in their propagation through the tidal fields of the cosmic large-scale structure \citep[see][as a review]{2001PhR...340..291B}. The lensing potential $\psi$ is a projection of the gravitational potential $\Phi$ along the line of sight, $\psi = 2\int\dd\chi\:W_\psi(\chi)\Phi$ with the weighting function $W_\psi(\chi)$,
\begin{equation}
W_\psi(\chi) = \frac{D_+(a)}{a}\frac{G(\chi)}{\chi}.
\end{equation}
$G(\chi)$ is the lensing-efficiency weighted galaxy redshift distribution,
\begin{equation}
G(\chi) = \int_\chi^{\chi_H}\dd\chip\: p(\chip) \frac{\dd z}{\dd\chip}\left(1-\frac{\chi}{\chip}\right)
\end{equation}
with $\dd z/\dd\chip = H(\chip) / c$. For the redshift distribution $p(z)\dd z$ we choose a standard parameterisation,
\begin{equation}
p(z)\dd z = p_0\left(\frac{z}{z_0}\right)^2\exp\left(-\left(\frac{z}{z_0}\right)^\beta\right)\dd z
\quad\mathrm{with}\quad 
\frac{1}{p_0}=\frac{z_0}{\beta}\Gamma\left(\frac{3}{\beta}\right),.
\end{equation}
The lensing observables follow from the lensing potential $\psi$ by taking second derivatives $\bpsi = \partial^2\psi/\partial\theta_i\partial\theta_j$ and contracting this tensor with the Pauli-matrices $\sigma_\alpha$ \citep{1972hmf..book.....A}. Specifically, the weak lensing convergence $\kappa$ is given by $\kappa = \trace(\bpsi\sigma_0)/2 = \Delta\psi/2$ and the two shear components $\gamma_+ = \trace(\bpsi\sigma_1)/2$, $\gamma_\times = \trace(\bpsi\sigma_3)/2$. Although the shear is the primary observable in weak lensing, we carry out our statistical investigations with the convergence as it has identical statistical properties and, being scalar, is easier to handle. For EUCLID, $z_0\simeq0.64$ such that the median redshift is 0.9.

\subsection{Polyspectra of the weak lensing convergence}
With the relation $\Delta\psi = 2\kappa$ is is straightforward to compute the angular spectrum $C_\kappa(\ell)$ of the weak lensing convergence from the spectrum $P_\Phi(k)$ of the gravitational potential,
\begin{equation}
C_\kappa(\ell) 
= \ell^4
\int_0^{\chi_H}\frac{\dd\chi}{\chi^2}\:W_\psi^2(\chi)P_\Phi(k)
\end{equation}
by application of the Limber-equation \citep{1954ApJ...119..655L}. Generalisation of the Limber-projection and repeated substitution of $\kappa = \ell^2\psi/2$ yields for the convergence bispectrum $B_\kappa(\vecl_1,\vecl_2,\vecl_3)$,
\begin{equation}
B_\kappa(\vecl_1,\vecl_2,\vecl_3) =
(\ell_1\ell_2\ell_3)^2
\int_0^{\chi_H}\frac{\dd\chi}{\chi^4}\:W_\psi^3(\chi)B_\Phi(\veck_1,\veck_2,\veck_3)
\end{equation}
and finally for the convergence trispectrum $T_\kappa(\vecl_1,\vecl_2,\vecl_3,\vecl_4)$,
\begin{equation}
T_\kappa(\vecl_1,\vecl_2,\vecl_3,\vecl_4)=
(\ell_1\ell_2\ell_3\ell_4)^2
\int_0^{\chi_H}\frac{\dd\chi}{\chi^6}\:W_\psi^4(\chi)T_\Phi(\veck_1,\veck_2,\veck_3,\veck_4).
\end{equation}
With these polyspectra it is then possible to derive cumulants of the convergence density field which can be smoothed on the angular scale $\theta$ by a function $W(\ell\theta)$, which we take to be Gaussian,
\begin{equation}
W(\ell\theta) = \exp\left(-\frac{(\ell\theta)^2}{2}\right).
\end{equation}
Consequently, the variance $\sigma^2$ of the smoothed convergence field reads
\begin{equation}
\kappa_2 = \sigma^2 = \int\frac{\ell\dd\ell}{2\pi}\:W(\ell\theta)^2\:C_\kappa(\ell),
\label{eqn_variance}
\end{equation}
which is equal to the second cumulant $\kappa_2$ of the distribution $p(\kappa)\dd\kappa$. The third cumulant $\kappa_3$ then follows from integration of the smoothed bispectrum \citep{Bernardeau:2002fk},
\begin{equation}
\kappa_3 = 
\int\frac{\dd^2\ell_1}{(2\pi)^2}W(\ell_1\theta)
\int\frac{\dd^2\ell_2}{(2\pi)^2}W(\ell_2\theta)
\int\frac{\dd^2\ell_3}{(2\pi)^2}W(\ell_3\theta)\:
B_\kappa(\vecl_1,\vecl_2,\vecl_3),
\label{eqn_skewness}
\end{equation}
and lastly, the fourth cumulant $\kappa_4$ can be obtained in complete analogy with
\begin{equation}
\kappa_4 =
\int\frac{\dd^2\ell_1}{(2\pi)^2}W(\ell_1\theta)
\int\frac{\dd^2\ell_2}{(2\pi)^2}W(\ell_2\theta)
\int\frac{\dd^2\ell_3}{(2\pi)^2}W(\ell_3\theta)
\int\frac{\dd^2\ell_4}{(2\pi)^2}W(\ell_4\theta)\:
T_\kappa(\vecl_1,\vecl_2,\vecl_3,\vecl_4).
\label{eqn_kurtosis}
\end{equation}

\begin{figure}
\begin{center}
\resizebox{0.5\hsize}{!}{\includegraphics{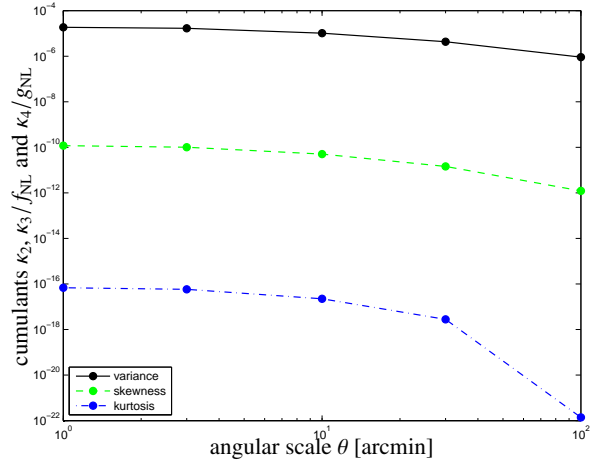}}
\caption{Cumulants $\kappa_2$, $\kappa_3/\fnl$ and $\kappa_4/\gnl$ as a function of angular scale $\theta$ for a Gaussian smoothing function $W(\ell\theta)$.}
\label{fig_cumulants}
\end{center}
\end{figure}

The Gaussian cumulant $\kappa_2=\sigma^2$, and the two non-Gaussian contributions $\kappa_3/\fnl$ and $\kappa_4/\gnl$ are depicted in Fig.~\ref{fig_cumulants} as a function of angular scale $\theta$. Quite generally, the two non-Gaussian cumulants will be proportional to the non-Gaussianity parameters $\fnl$ and $\gnl$, and all cumulants are decreasing with smoothing scale, because the fluctuations are wiped out and the cumulants as an integrated measure of the fluctuation amplitude decrease. As emphasised by \citet{Jeong:2011uq}, the non-Gaussianity in the observable is weakened due to the central limit theorem because in the line of sight integration many non-Gaussian values for the gravitaitonal potential are added that yield an approximately Gaussian result.

The cumulant $\kappa_4$ is very small for $\tnl$-type non-Gaussianity, about three orders of magnitude less relative to that generated by $\gnl$, which is the reason why we do not include it in the subsequent calculations. The reason of this behaviour derives from the fact that the weigthing functions $W(\ell_i\theta)$ downweight contributions from large multipoles $\ell_i$. The integrations in eqns.~(\ref{eqn_skewness}) and~(\ref{eqn_kurtosis}) needed for the cumulants $\kappa_3$ and $\kappa_4$ are carried out in polar coordinates with a Monte-Carlo scheme \citep[specifically, with the CUBA-library by][who provides a range of adaptive Monte-Carlo integration algorithms]{Hahn:2005uq}, which reduces the computational complexity considerably.

\section{Extreme value statistics}\label{sect_extreme}

\subsection{Gram-Charlier series}
If a Gaussian distribution with zero mean and variance $\sigma^2=\kappa_2$ is weakly perturbed by the presence of a non-vanishing third and fourth cumulant $\kappa_3$ and $\kappa_4$, respectively, the distribution $p(\kappa)\dd\kappa$ can be approximated with the Gram-Charlier-series \citep[see][who in addition quantify the limits of applicability of the expansion]{wallace, greenwood_durand1, greenwood_durand2, Colombi:1994vn, Juszkiewicz:1995kx, Bernardeau:2002fk},
\begin{equation}
p(\kappa)\dd\kappa = 
\frac{1}{\sqrt{2\pi\sigma^2}}\exp\left(-\frac{\kappa^2}{2\sigma^2}\right)\times
\left[
1 + 
\frac{\kappa_3}{3!\sigma^3}H_3\left(\frac{\kappa}{\sigma}\right) + 
\frac{\kappa_4}{4!\sigma^4}H_4\left(\frac{\kappa}{\sigma}\right)
\right]\dd\kappa
\label{eqn_gram_charlier}
\end{equation}
with the argument $x = \kappa/\sigma$ of the Hermite polynomials $H_n(x)$, which can be computed by $n$-fold differentiation of a Gaussian,
\begin{equation}
H_n\left(\frac{\kappa}{\sigma}\right) = 
(-\sigma)^n\exp\left(\frac{\kappa^2}{2\sigma^2}\right)\:\frac{\dd^n}{\dd \kappa^n}\:\exp\left(-\frac{\kappa^2}{2\sigma^2}\right).
\label{eqn_hermite}
\end{equation}
It is worth noting that the perturbation of $p(\kappa)\dd\kappa$ with $H_3$ and $H_4$ do not change the mean and the variance. Specifically, the Hermite-polynomials needed read \citep{1972hmf..book.....A}:
\begin{equation}
H_1(x) = x
,\quad
H_2(x) = x^2-1
,\quad
H_3(x) = x^3-3x
,\quad
H_4(x) = x^4 - 6x^2 + 3
\quad\mathrm{and~later}\quad
H_5(x) = x^5 - 10x^3 + 15x.
\end{equation}
By substituting eqn.~(\ref{eqn_hermite}) and integrating by parts the cumulative function $P(\kappa)$ of the Gram-Charlier-distribution $p(\kappa)\dd\kappa$ can be written down analytically,
\begin{equation}
P(\kappa) = \int_{-\infty}^\kappa\dd\kappa^\prime\:p(\kappa^\prime) = 
\Phi\left(\frac{\kappa}{\sigma}\right) -
\frac{1}{\sqrt{2\pi\sigma^2}}\exp\left(-\frac{\kappa^2}{2\sigma^2}\right)
\left[
\frac{\kappa_3}{3!\sigma^2}H_2\left(\frac{\kappa}{\sigma}\right)+
\frac{\kappa_4}{4!\sigma^3}H_3\left(\frac{\kappa}{\sigma}\right)
\right]
\label{eqn_cumulative}
\end{equation}
where the cumulative function $\Phi(\kappa/\sigma)$ of the Gaussian distribution is expressed in terms of the error function $\mathrm{erf}(\kappa/\sigma)$,
\begin{equation}
\Phi\left(\frac{\kappa}{\sigma}\right) = 
\frac{1}{2}\left(1 + \mathrm{erf}\left(\frac{\kappa}{\sqrt{2}\sigma}\right)\right),
\end{equation}
as defined by \citet{1972hmf..book.....A}. By using the derivative relation 
\begin{equation}
\frac{\dd}{\dd\kappa} H_n\left(\frac{\kappa}{\sigma}\right) = 
\frac{n}{\sigma}\:H_{n-1}\left(\frac{\kappa}{\sigma}\right)
\end{equation}
of the Hermite polynomials $H_n(x)$, the derivative of the Gram-Charlier distribution takes the compact analytical form,
\begin{equation}
\frac{\dd}{\dd\kappa}p(\kappa) = 
-\frac{1}{\sqrt{2\pi\sigma^2}}\exp\left(-\frac{\kappa^2}{2\sigma^2}\right)\times
\left[
\frac{\kappa}{\sigma^2} +
\frac{\kappa_3}{3!\sigma^4}H_4\left(\frac{\kappa}{\sigma}\right) + 
\frac{\kappa_4}{4!\sigma^5}H_5\left(\frac{\kappa}{\sigma}\right)
\right].
\label{eqn_derivative}
\end{equation}
The moment generating function $M(k)$ can be computed analytically as well,
\begin{equation}
M(k) 
= \int\dd\kappa\:\exp(k\kappa)p(\kappa)
= \exp\left( \frac{\sigma^2 k^2}{2} \right) \times 
\left[ 1 + \frac{\kappa_3}{3!}k^3 +  \frac{\kappa_4}{4!}k^4 \right],
\end{equation}
from which the moments of order $n$ can be obtained by $n$-fold differentiation and setting $k$ to zero.
We would like to add that the Gram-Charlier expansion in eqn.~(\ref{eqn_gram_charlier}) is only applicable for weak non-Gaussianities in which $\kappa_3\ll\sigma^3$ and $\kappa_4\ll\sigma^4$, because otherwise the Hermite-polynomials could cause negative values for $p(\kappa)\dd\kappa$. The regime of weak non-Gaussianity in the weak lensing signal would be left if $\fnl\gsim10^4$ and $\gnl\gsim10^8$, depending on angular scale.

\subsection{Number of samples}
\begin{figure}
\begin{center}
\resizebox{0.5\hsize}{!}{\includegraphics{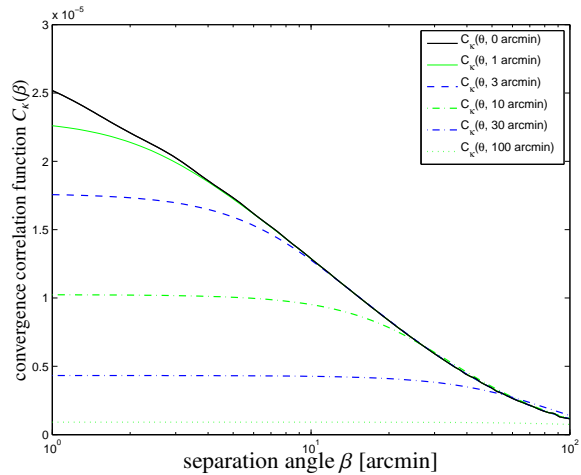}}
\caption{The angular correlation function $C_\kappa(\beta)$ is shown as a function of separation angle $\beta$ (black solid line) and for a range of smoothing scales, $\theta = 1,3,10,30,100$ arcmin, for a Gaussian filter.}
\label{fig_correlation}
\end{center}
\end{figure}

\begin{figure}
\begin{center}
\resizebox{0.5\hsize}{!}{\includegraphics{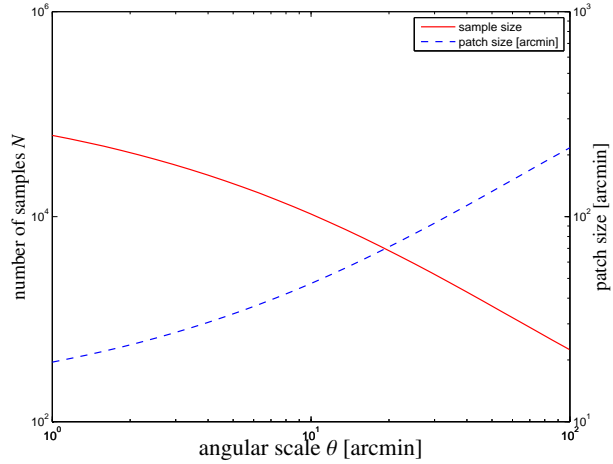}}
\caption{Sample size $N$ as a function of smoothing scale $\theta$ employed in calculating the cumulants $\kappa_n$, $n=2,3,4$. The sample size is computed from the angular scale at which the correlation function drops to a fraction of $\exp(-1)$ of its value at zero lag.}
\label{fig_samples}
\end{center}
\end{figure}

We compute an estimate of the number $N$ of samples from the correlation function $C_\kappa(\beta)$ of the convergence field $\kappa$ that has been smoothed on the scale $\theta$,
\begin{equation}
C_\kappa(\beta) = \int\frac{\ell\dd\ell}{2\pi}\:W(\ell\theta)^2C_\kappa(\ell)\:J_0(\ell\beta),
\end{equation}
which is depicted in Fig.~\ref{fig_correlation} for a range of smoothing scales and for a Gaussian window function $W(\ell\theta)=\exp(-(\ell\theta)/2)$. The correlation function allows us to define a correlation length $\beta$ at which the value of $C_\kappa(\beta)$ has dropped to a fraction $\exp(-1)$ of its value at zero lag, $C_\kappa(\beta=0) = \sigma^2 = \kappa_2$. The number of samples $N$ can then be estimated with the relation $N\times\pi\beta^2=4\pi f_\mathrm{sky}$, i.e. the number of patches of area $\pi\beta^2$ that could be fitted in the survey solid angle. In this approximated picture, the smoothed random field is taken to assume independent values $\kappa$ in patches of size $\beta$. The number of available samples $N$ as a function of smoothing scale $\theta$ is given in Fig.~\ref{fig_samples}, where we consider the case of the EUCLID mission with the sky fraction $f_\mathrm{sky}=1/2$: $N$ drops from $\simeq 4\times10^4$ if there is hardly any smoothing at $\theta=1~\mathrm{arcmin}$ to a few hundred if a strong smoothing on the scale $\theta=100~\mathrm{arcmin}$ is applied. 

We choose the smoothing scale $\theta=10~\mathrm{arcmin}$ for the subsequent analysis in order to have sufficiently interesting sample sizes while avoiding a possible strong contamination from non-Gaussianities that evolve in nonlinear structure formation. Fig.~\ref{fig_spectrum} compares smoothed convergence spectra resulting from linear and nonlinear CDM-spectra and we found a contamination of the variance $\sigma^2$ amounting to $~\simeq 7\%$ at $\theta=10~\mathrm{arcmin}$, compared to $\simeq 1\%$ at $\theta=30~\mathrm{arcmin}$ and $\simeq 18\%$ at $\theta=3~\mathrm{arcmin}$.

We would like to point out that in drawing extreme values it would be incorrect to generate a vector of $N$ random deviates for $\kappa$ and identify in this vector the largest and smallest sample. Instead, one needs to carry out the numerical experiment for finding the largest and the smallest sample separately. The reason for this is the fact that samples for extreme values are compared to each other for finding the extrema, and for this process $N$ samples are needed, which must not be reused as would be the case in the first approach: Each time a new value is drawn, it must be given the chance (and hence probability) of being larger than the current maximum but at the same time smaller than the current minimum. Therefore, every time a new value is drawn, the comparison with the current largest value and the comparison with the smallest one must be separate processes. The importance of this separation can be clearly seen when considering the very first value which is drawn, since this value is at the same time the largest and the smallest one. This actually also reduces the effective number of samples by 1.

\subsection{Sampling from the Gram-Charlier distribution}
With the analytical form eqn.~(\ref{eqn_cumulative}) of the cumulative distribution $P(\kappa)$ it is possible to sample from the Gram-Charlier-distribution $p(\kappa)\dd\kappa$ by using its invertibility: From a sample $y$ of the uniform distribution from the unit interval one can obtain a sample of $\kappa$ by setting $\kappa = P^{-1}(y)$. This inverse always exists because $P(\kappa)$ as an integral of a positive function is monotonically increasing and therefore invertible. Likewise, samples from the extreme value distributions can be generated by drawing $N$ random numbers from the uniform distribution, and by mapping the largest (and the smallest) of those samples onto $\kappa$. Because the cumulative distribution $P(\kappa)$ is monotonic, the largest sample of $y$ will be converted to the largest value in $\kappa$, and likewise the smallest sample of $y$ will be the smallest $\kappa$-value. This approach has advantages over direct sampling from the Gram-Charlier distribution and finding the extrema, because the inversion $y\rightarrow\kappa$ has to be carried out only once.

\subsection{Extreme value distributions}
The reasoning behind extreme value distributions is very instructive \citep{Gumbel:1954kx,Gumbel:2004vn}: The cumulative distribution $P(\kappa)$ gives the probability that a sample is drawn with a value $<\kappa$, and consequently $P(\kappa)^N$ indicates the probability that $N$ independent samples are all smaller than $\kappa$. The probability of the complementary event, i.e. that at least a single one of the samples is larger than $\kappa$ would then be given by $P_+(\kappa) = 1-P(\kappa)^N$. Differentiation yields the distribution $p_+(\kappa)\dd\kappa$ of the maximum values drawn from $p(\kappa)\dd\kappa$ in $N$ trials:
\begin{equation}
p_+(\kappa) = \frac{\dd}{\dd\kappa}P_+(\kappa) = NP(\kappa)^{N-1}p(\kappa),
\end{equation}
which can be computed analytically with eqns.~(\ref{eqn_gram_charlier}) and~(\ref{eqn_cumulative}).
The argumentation for the smallest samples proceeds in complete analogy: Again, the cumulative distribution $1-P(\kappa)$ states the probability that a sample is $>\kappa$, and the probability that $N$ independent samples are all larger than $\kappa$ would be given by $(1-P(\kappa))^N$. The complementary case of a single sample being smaller than $\kappa$ is computed with $P_-(\kappa) = 1-(1-P(\kappa))^N$, which can be differentiated to get the extreme value distribution $p_-(\kappa)\dd\kappa$ of the minimum obtained in $N$ draws,
\begin{equation}
p_-(\kappa) = \frac{\dd}{\dd\kappa}P_-(\kappa) = N(1-P(\kappa))^{N-1}p(\kappa)
\end{equation}
By the derivation of the extreme value distribution $p_\pm(\kappa)\dd\kappa$ as the $N$-fold exponentiation of the cumulative function $P(\kappa)$ the distribution acquires naturally a strong sensitivity on the asymptotic behaviour of the distribution $p(\kappa)\dd\kappa$. In our case, the distribution will be influenced by the presence of a non-vanishing third and fourth cumulant are sourced by the three lowest-order inflationary non-Gaussianity parameters $\fnl$ and $\gnl$. Local non-Gaussianity from the $\tnl$-term influences $\kappa_4$ only weakly and will be neglected in the analysis.

\begin{figure}
\begin{center}
\resizebox{0.5\hsize}{!}{\includegraphics{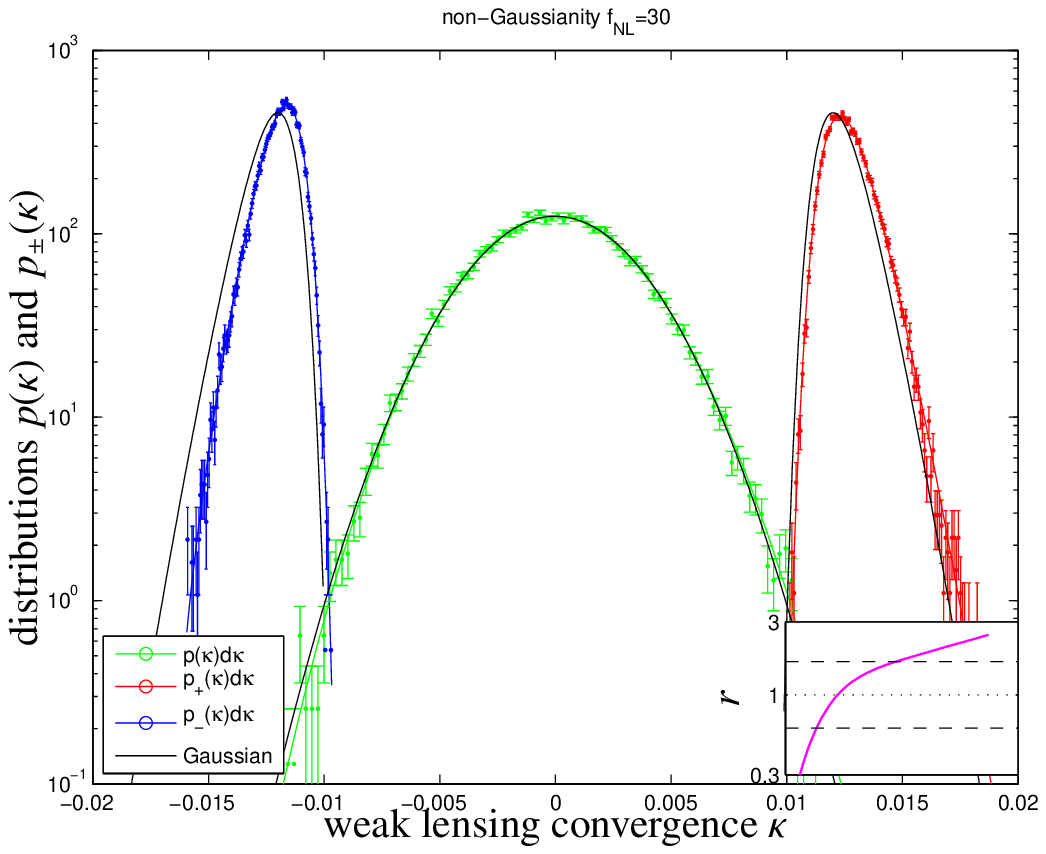}}
\caption{Gram-Charlier distribution $p(\kappa)\dd\kappa$ and the two extreme value distributions $p_\pm(\kappa)\dd\kappa$ with the non-Gaussianity parameters $\fnl=30$ and $\gnl=0$ on the angular scale $\theta=10$~arcmin, with a yield of $N=10597$ samples. Additionally, we show show samples from the Gram-Charlier distribution $p(\kappa)\dd\kappa$ including Poissonian errors and the two extreme value distributions $p_\pm(\kappa)\dd\kappa$ for the Gaussian reference model. The inset figure shows the ratio of the extreme value distributions $p_+(\kappa)\dd\kappa$ between the Gaussian and the non-Gaussian model, with varying $\kappa$ along with lines marking the ratios $\exp(\pm1/2)$ indicating an equivalent $1\sigma$ change in likelihood.}
\label{fig_gram_charlier_fnl}
\end{center}
\end{figure}

The Gram-Charlier distribution $p(\kappa)\dd\kappa$ along with the two extreme value distributions $p_\pm(\kappa)\dd\kappa$ are shown for $\fnl=30$ and for $\theta=10~\mathrm{arcmin}$ (corresponding to $N=10597$ on EUCLID's survey cone) in Fig.~\ref{fig_gram_charlier_fnl}. While there is a very small asymmetry in the distribution $p(\kappa)\dd\kappa$ of the convergences themselves, the skewness introduced by $\fnl$ gives rise to a much larger asymmetry in the extreme value distributions $p_\pm(\kappa)\dd\kappa$. Positive $\fnl$ skew the distribution in the direction of positive values, making large maxima more likely and large minima less likely. The samples for the Gram-Charlier distribution and the direct sampling of the extreme value distributions corresponds very well to the analytical expressions. Even without the influence of non-Gaussianities one sees that values as large as $\kappa=0.012$ are the most likely to be expected for the sample size, corresponding to random events at a distance of $\simeq3.4\sigma$ away from the mean at zero. Extreme value of that magnitude are consistent with the fact that with $N\simeq10^4$ samples it is possible to probe the wings of the Gaussian distribution at probabilities of $\mathrm{erfc}(3.4/\sqrt{2})\simeq6\times10^{-4}$.

\begin{figure}
\begin{center}
\resizebox{0.5\hsize}{!}{\includegraphics{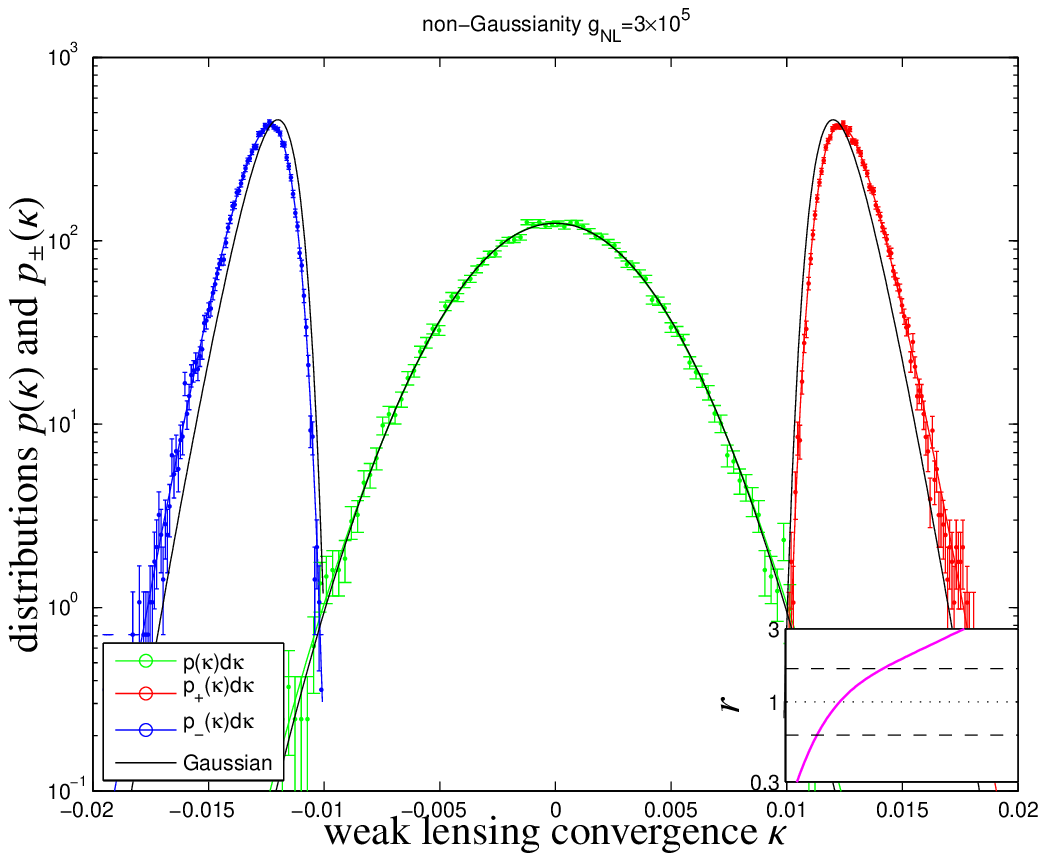}}
\caption{Gram-Charlier distribution $p(\kappa)\dd\kappa$ and the two extreme value distributions $p_\pm(\kappa)\dd\kappa$ with the non-Gaussianity parameters $\gnl=3\times10^5$ and $\fnl=0$ on the angular scale $\theta=10$~arcmin, yielding $N=10597$ samples. Additionally, we show show samples from the Gram-Charlier distribution $p(\kappa)\dd\kappa$ including Poissonian errors and the two extreme value distributions $p_\pm(\kappa)\dd\kappa$ for the Gaussian reference model. The inset gives the ratio between the extreme value distributions $p_+(\kappa)\dd\kappa$ for the Gaussian and the non-Gaussian parent distribution $p(\kappa)\dd\kappa$, with lines indicating the ratios $\exp(\pm 1/2)$, which corresponds to a $1\sigma$ change in likelihood.}
\label{fig_gram_charlier_gnl}
\end{center}
\end{figure}

Fig.~\ref{fig_gram_charlier_gnl} shows the Gram-Charlier distribution $p(\kappa)\dd\kappa$ with the two corresponding extreme value distributions $p_\pm(\kappa)\dd\kappa$ with $\gnl=3\times10^5$ and on an angular scale $\theta=10~\mathrm{arcmin}$. Introducing a positive kurtosis into the distribution is difficult to see in the distribution itself, but makes large extremes much more likely. Overall, the sensitivity of the extreme values to a non-Gaussian kurtosis is much weaker compared to that of a non-Gaussian skewness, and there is a very good correspondence between the sampled distributions and the analytical expressions.

\subsection{Posterior statistics of the Gram-Charlier distribution}\label{sect_posterior}
In this section we investigate the properties of the extreme value distributions $p_\pm(\kappa)\dd\kappa$ in more detail by deriving its average, its most likely value and its median and by relating its first moments to the standard parameters of the Gumbel distribution. We focus on particular on the position of the extreme value distribution as a function of smoothing $\theta$ which influences both the magnitudes of the cumulants $\kappa_n$ as well as the number of samples $N$, which is the dominating quantity. As seen in the two previous plots, the extreme value statistics generates a much stronger difference between extreme values from small differences in the parent distributions.

The average $\bar{\kappa}_\pm$ of the extreme value distribution $p_\pm(\kappa)\dd\kappa$ is given by 
\begin{equation}
\bar{\kappa}_\pm = \int\dd\kappa\:\kappa p_\pm(\kappa).
\end{equation}
The most likely value $\hat{\kappa}_\pm$ follows from solving:
\begin{equation}
\frac{\dd}{\dd\kappa}p_\pm(\kappa) = 0,
\label{eqn_most_likely}
\end{equation}
where the analytical form eqn.~(\ref{eqn_derivative}) of the derivative $\dd p(\kappa)/\dd\kappa$ is particularly useful. Likewise, the median $\tilde{\kappa}_\pm$ can be computed by solving 
\begin{equation}
P_\pm(\kappa) = \frac{1}{2}.
\end{equation}

\begin{figure}
\begin{center}
\resizebox{0.5\hsize}{!}{\includegraphics{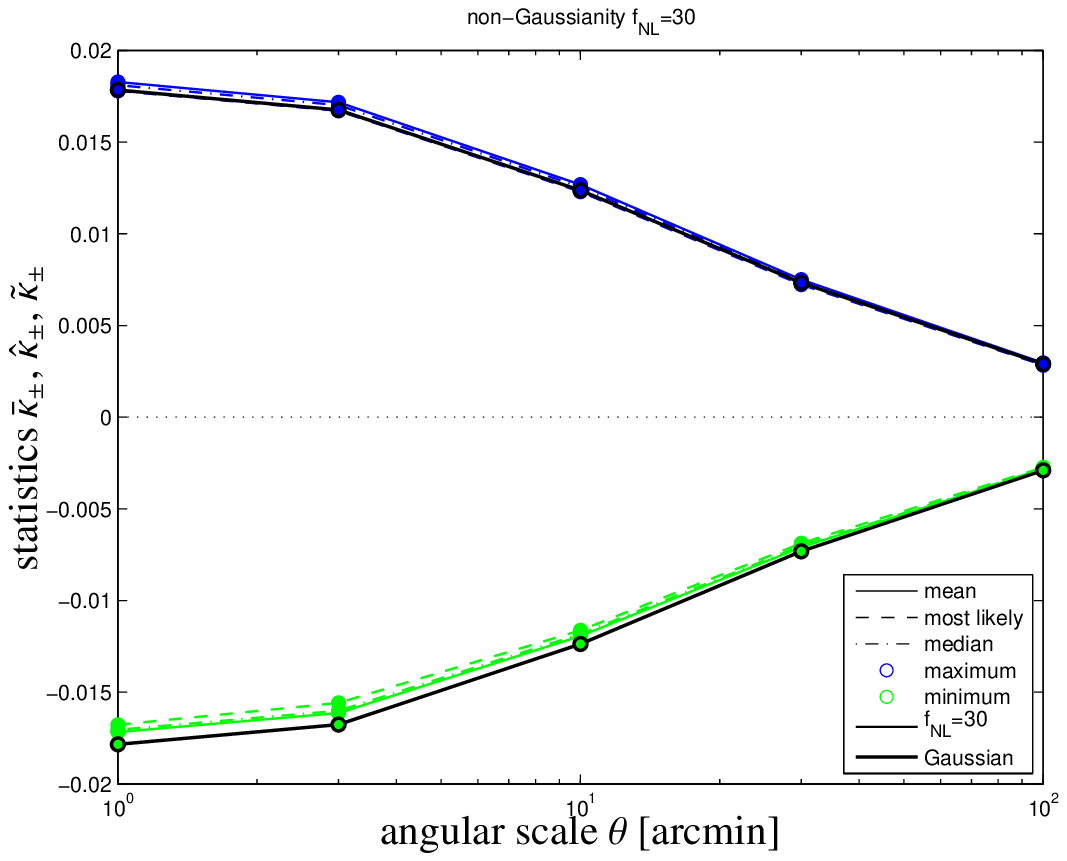}}
\caption{The mean value $\bar{\kappa}_\pm$, the most likely value $\hat{\kappa}_\pm$ and the median $\tilde{\kappa}_\pm$ of the extreme value distribution $p_\pm(\kappa)\dd\kappa$ for a non-Gaussian model with $\fnl=30$ and $\gnl=0$ in comparison to a Gaussian model, as a function of angular scale $\theta$.}
\label{fig_post_fnl}
\end{center}
\end{figure}

\begin{figure}
\begin{center}
\resizebox{0.5\hsize}{!}{\includegraphics{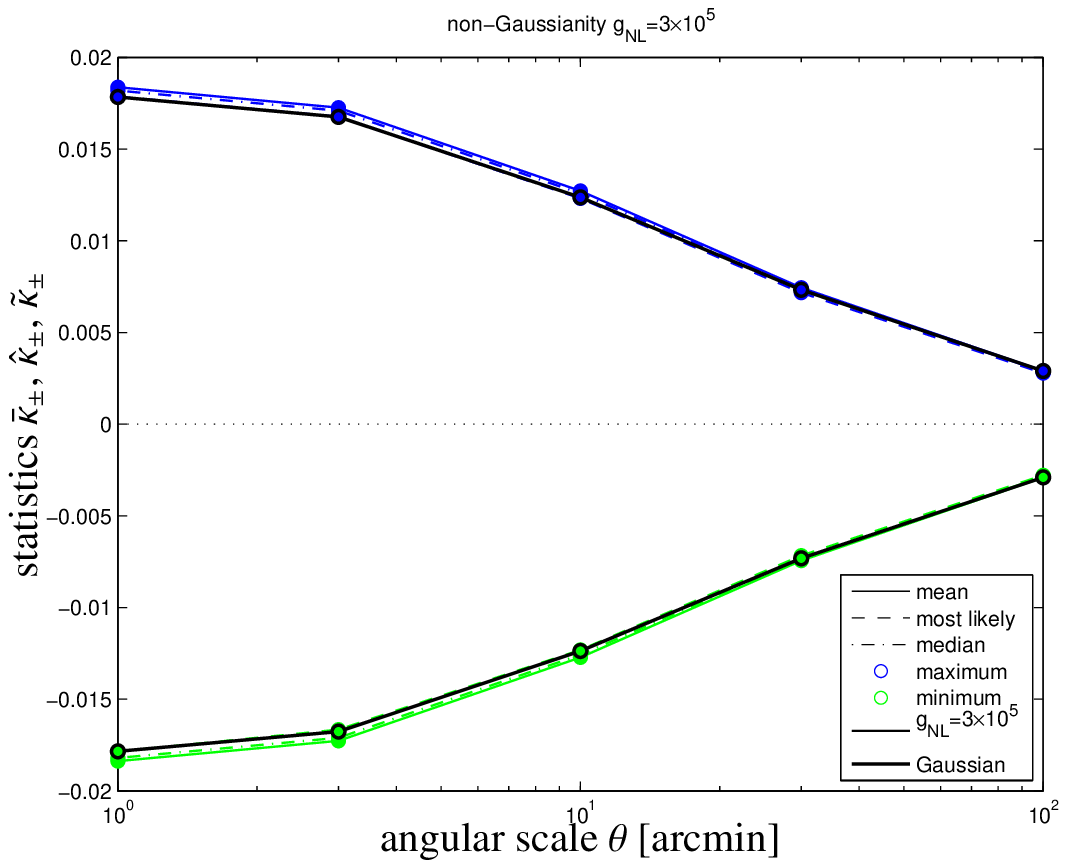}}
\caption{The median value $\bar{\kappa}_\pm$, the most likely value $\hat{\kappa}_\pm$ and the median $\tilde{\kappa}_\pm$ of the extreme value distribution $p_\pm(\kappa)\dd\kappa$ for a non-Gaussian model with $\gnl=3\times10^5$ and $\fnl=0$ in comparison to a Gaussian model, as a function of angular smoothing scale $\theta$.}
\label{fig_post_gnl}
\end{center}
\end{figure}

Figs.~\ref{fig_post_fnl} and~\ref{fig_post_gnl} give an impression of how fast the extreme value distribution shifts away from the parent distribution if the smoothing scale $\theta$ is varied, due to changes in the cumulants $\kappa_n$ and the number of available samples $N$, the latter being the driving factor, as mentioned previously. As already apparent from Figs.~\ref{fig_gram_charlier_fnl} and~\ref{fig_gram_charlier_gnl}, a nonzero positive $\fnl$ skews the distribution and shifts the maximum distribution $p_+(\kappa)\dd\kappa$ towards larger values and the minimum distribution $p_-(\kappa\dd\kappa)$ towards less negative values. Non-zero $\gnl$ causes larger absolute values for both $p_+(\kappa)\dd\kappa$ and $p_-(\kappa)\dd\kappa$. As expected for a unimodal distribution, the means $\bar{\kappa}_\pm$, the most likely values $\hat{\kappa}_\pm$ and the median values $\tilde{\kappa}_\pm$ show a very similar behaviour. In both cases, the position of the extreme value distribution tends towards zero with increasing smoothing scale $\theta$, which reduces the sample number $N$ as well as the numerical value of all cumulants.

\subsection{Relation to the Gumbel-distribution}
The parameters $\mu$ and $\beta$ of the standard Gumbel distribution can be derived from the mean and the variance of $p_\pm(\kappa)\dd\kappa$,
\begin{equation}
\frac{\beta^2\pi^2}{6} = \int\dd\kappa\:\kappa^2p_\pm(\kappa)
\quad\mathrm{and}\quad
\mu + \gamma\beta = \int\dd\kappa\:\kappa p_\pm(\kappa).
\end{equation}
with the Euler-Mascheroni-constant, $\gamma\simeq0.57721$ \citep{1972hmf..book.....A}. One naturally recovers the shape of the Gumbel distribution in the limit of large $N$ which can be seen from the cumulative distribution $P_+(\kappa) = P^N(\kappa) = \exp\left(N\ln P(\kappa)\right) = \exp\left(N\ln\left(1-(1-P(\kappa))\right)\right)\simeq \exp\left(-N(1-P(\kappa))\right)$ applying a Taylor expansion of the logarithm in the last step. Substituting the Gaussian distribution $p(\kappa)/\kappa$ as a approximation for $1-P(\kappa)$ for large $\kappa$ one obtains the Gumbel distribution $P_+(\kappa)\simeq\exp(-N/\kappa\:\exp(-\kappa^2/(2\sigma^2)))$ \citep{Gumbel:2004vn}.

\begin{figure}
\begin{center}
\resizebox{0.5\hsize}{!}{\includegraphics{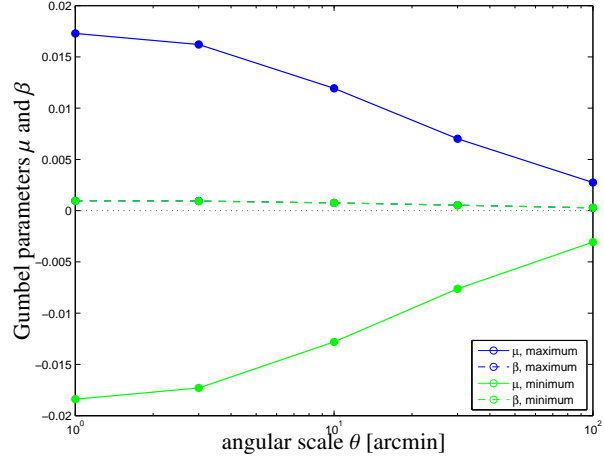}}
\caption{Gumbel parameters $\mu$ and $\beta$ for the extreme value distributions $p_\pm(\kappa)\dd\kappa$ resulting from a Gaussian parent distribution $p(\kappa)\dd\kappa$, as a function of angular scale $\theta$.}
\label{fig_gumbel}
\end{center}
\end{figure}

Fig.~\ref{fig_gumbel} illustrates the variation of the two parameters $\mu$ and $\beta$ with angular scale if the Gaussian distribution is approximated with an extreme value distribution of the Gumbel-shape. Clearly, the position of the mean value distribution described by $\mu$ decreases if the sample number and the variance of the parent distribution decrease, and the same argument applies to the width of the extreme value distribution. Because the perturbation with Hermite polynomials in the Gram-Charlier distribution does not introduce a different asymptotic behaviour than that of a Gaussian distribution, the extreme value distribution is of approximate Gumbel-shape and weak non-Gaussianities do not affect the general shape of the extreme value distribution.

\subsection{Inference from extreme values}
Although extreme value statistics seems to be applicable in situations where models are excluded because they might be implausible in generating a certain observed extreme value, they can in fact it can be used for parameter inference, e.g. for the non-Gaussianity parameters $\fnl$ and $\gnl$: When observing a certain extreme value $\kappa$, one can consider the distribution $p_\pm(\kappa|\fnl)$ with its dependence on the parameter set $\fnl$ as a likelihood, and compare different likelihoods by their ratio $r$, 
\begin{equation}
r(\kappa,\fnl) = \frac{p_\pm(\kappa|\fnl)}{p_\pm(\kappa|\fnl=0)}
\quad\mathrm{or}\quad
r(\kappa,\gnl) = \frac{p_\pm(\kappa|\gnl)}{p_\pm(\kappa|\gnl=0)},
\end{equation}
which, according to the Neyman-Pearson lemma, is the most effective test for distinguishing the likelihoods that certain parameter choices provide an explanation of the data, i.e. the observed extreme value $\kappa$ in our case. The likelihood ratios $r(\fnl)$ and $r(\gnl)$ in our example would quantify the plausibility of a cosmological model with nonzero non-Gaussianities $\fnl$ or $\gnl$ relative to a purely Gaussian fiducial model in providing an explanation to an observed extreme value.

The insets in Figs.~\ref{fig_gram_charlier_fnl} and~\ref{fig_gram_charlier_gnl} show the likelihood ratios $r$ as a function of $\kappa$ between the extreme value distributions from the non-Gaussian and the Gaussian model. For weak non-Gaussianities, the likelihood ratio $r$ as a function of $\kappa$ assumes values close to unity if the extreme sample is close to the most likely sample for a particular Gram-Charlier-distribution but would assumes values differing significantly from one if the sample is much larger or smaller than the most likely value.

\begin{figure}
\begin{center}
\resizebox{0.5\hsize}{!}{\includegraphics{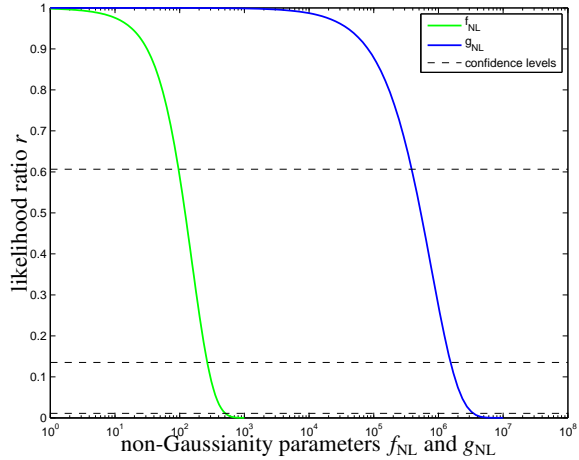}}
\caption{Likelihood ratios $r$ for varying $\fnl$ (green line) and $\gnl$ (blue line) evaluated at the most likely maximum value obtained in the Gaussian reference model with $\fnl=\gnl=0$. The angular smoothing scale is set to $10~\mathrm{arcmin}$, resulting in $N=10597$ samples. The horizontal lines indicate confidence levels corresponding to $n\sigma$, $n=1,2,3$.}
\label{fig_fit}
\end{center}
\end{figure}

Fig.~\ref{fig_fit} shows the likelihood ratios $r$ as a function of either $\fnl$ or $\gnl$ if the reference model is Gaussian with $\fnl=\gnl=0$. We choose to evaluate the likelihood ratio for a value of $\kappa$ that occurs in the random experiment with the highest probability, i.e. the most likely value $\hat{\kappa}$ derived with eqn.~(\ref{eqn_most_likely}) for $p_+(\kappa)\dd\kappa$. We focus on the maximum value of the convergence $\hat{\kappa}_+$ which can be computed analytically for the Gram-Charlier-distribution. The choice of $\hat{\kappa}_\pm$ for estimating the width of the likelihood corresponds to the average $\bra\chi^2\ket$ of the $\chi^2$-functional in conventional fits for unbiased models. In this sense, we are attempting to carry out a fit with a single measurement and estimate the precision of the parameter inference from that single measurement. Fig.~\ref{fig_fit} suggests constraints of the order $\Delta\fnl\simeq10^2$ and $\Delta\gnl\simeq10^5$ from extreme value statistics, i.e. an observation of the single extreme value for such a non-Gaussianity would be incompatible with a Gaussian parent distribution.

We conclude similarly to \citet{Mikelsons:2009dz,Chongchitnan:2012oq} that the extreme values are not competitive in their sensitivity to weak non-Gaussianities, at least for typical extrema, even though the simplicity of the measurement could be attractive. While extreme values of the lensing convergence might provide a consistency check for constraints on $\fnl$, their very weak sensitivity on $\gnl$ makes it doubtful if meaningful constraints on primordial trispectra can be derived from extreme value statistics, even less so for $\tnl$-type non-Gaussianity. By running a direct estimate of the primordial bispectrum in a non-tomographic setup very similar constraints on $\fnl$ of $\sim10^2$ are within reach with EUCLID \citep{Schafer:2012fk}, and corresponding constraints on $\gnl$ are of the order of $\sim10^5$, while these values can be improved substantially by lensing tomography. In comparison, large-scale structure probes other than lensing are able to provide constraints close to order unity on $\fnl$, likewise the cosmic microwave background.

\section{Summary}\label{sect_summary}
Subject of this paper is the extreme value statistics of the weak lensing convergence in the presence of primordial inflationary non-Gaussianities. We would like to answer the question if the most extreme values of the weak lensing convergence averaged in apertures of a certain angular size is indicative of the non-Gaussianity parameters $\fnl$ and $\gnl$ in a basic local non-Gaussianity model.

\begin{enumerate}
\item{For this purpose, we perturb a Gaussian distribution for the lensing convergence with Hermite polynomials whose amplitudes are the cumulants of third and fourth order, i.e. with a Gram-Charlier series. These two cumulants are proportional to the parameters $\fnl$ and $\gnl$ and are computed from the local non-Gaussianity bi- and trispectra in a configuration space integration for which we use a very efficient adaptive Monte-Carlo integration. For investigating the dependence on angular scale, we introduce a Gaussian smoothing into the polyspectra and we find all cumulants to be decreasing functions with smoothign scale. We made sure that the smoothing is sufficiently strong such that small-scale structure formation non-Gaussianities have a small impact on the cumulants. The $\tnl$-term provides a much smaller contribution to the weak lensing trispectrum in comparison to the $\gnl$-part and for that reason we neglect it in our investigation.}
\item{The Gram-Charlier distribution has the convenient property of analytical expressions for the cumulative distribution, the derivative of the distribution and the moment-generating function. We provide analytical expressions for the extreme value distributions for drawing $N$ samples, which alleviates the usage of the generic Gumbel-distribution which would be recovered in the limit of large $N$. In EUCLID's weak lensing survey one can expect individual extreme values of the weak lensing convergence of a percent on the scale $\theta=10~\mathrm{arcmin}$. If Gaussian statistics is assumed, the most likely extreme value differs from the mean by $\simeq3.4\sigma$.}
\item{We propose an efficient sampling scheme for drawing Gram-Charlier distributed random numbers based on drawing uniformly distributed numbers from the unit interval and determining the extremes of this distribution before mapping it onto the weak lensing convergence with the inverse of the cumulative distribution $P(\kappa)$. We verified the correspondence between analytical results and samples from the extreme value distribution and found excellent agreement. The number of samples is estimated from the correlation length of the random field, where we make estimate the correlation length of the field by requiring that the correlation function has dropped to a fraction of $\exp(-1)$ from its value at zero lag and by tiling the survey area with circular patches of this size.}
\item{We investigated the sensitivity of extreme value distributions on constraining inflationary non-Gaussianity parameters. While non-Gaussianities change the parent distribution only weakly, the difference  between a non-Gaussian and a Gaussian model is amplified in the extreme value distribution.}
\item{We characterised the extreme value distribution and related it to the generic shape of the Gumbel-distribution, which is always recovered in the case of large sample numbers for a unrestricted random process. The mean value, the most likely value and the median of the extreme value distribution reflect the non-Gaussianity in the parent distribution and decrease with stronger smoothing because of two reasons: firstly, the cumulants decrease in value of a stronger smoothing is applied, and secondly, the number of available samples decreases because the correlation length of the convergence field increases.}
\item{By considering the likelihood ratio between the hypothesis that a non-Gaussian distribution provides and explanation of an extreme value compared to the null-hypothesis of a Gaussian parent distribution we show that individual extreme values can provide constraints on $\fnl$ of the order $10^2$ and on $\gnl$ of the order $10^5$. One can expect a significant improvement in these constraints if the sequence of the $n$th largest extrema is considered, similarly to \citet{Waizmann:2012kx} for the observation of massive clusters of galaxies. Due to the smallness of the contribution of the $\tnl$-term to the fourth cumulant $\kappa_4$ we did not derive a limit on $\tnl$.}
\end{enumerate}

In summary we would like to point out the simplicity of the statistical inference from weak lensing extreme values. We are in the process of extending our studies for the related case of structure formation non-Gaussianities, where an effective description of the convergence field with the lognormal distribution is applicable, and to the case of non-zero covariances between individual samples \citep[as an application of the formalism by][]{Bertin:2006fv}.

\section*{Acknowledgements}
FC receives funding through GSFP/Heidelberg and DFG's SPP1177, AFK acknowledges funding from DFG and BMS's work was supported by DFG within the framework of the excellence initiative through the Heidelberg Graduate School of Fundamental Physics. FC and AFK acknowledge support from the Graduate School for Fundamental Physics and the International Max-Planck Research School of Astronomy and Cosmic Physics. We would like to thank Chris Byrnes and Jean-Claude Waizmann for their thoughtful comments and suggestions.

\bibliography{bibtex/aamnem,bibtex/references}
\bibliographystyle{mn2e}

\appendix

\section{smoothed convergence spectra}
\begin{figure}
\begin{center}
\resizebox{0.5\hsize}{!}{\includegraphics{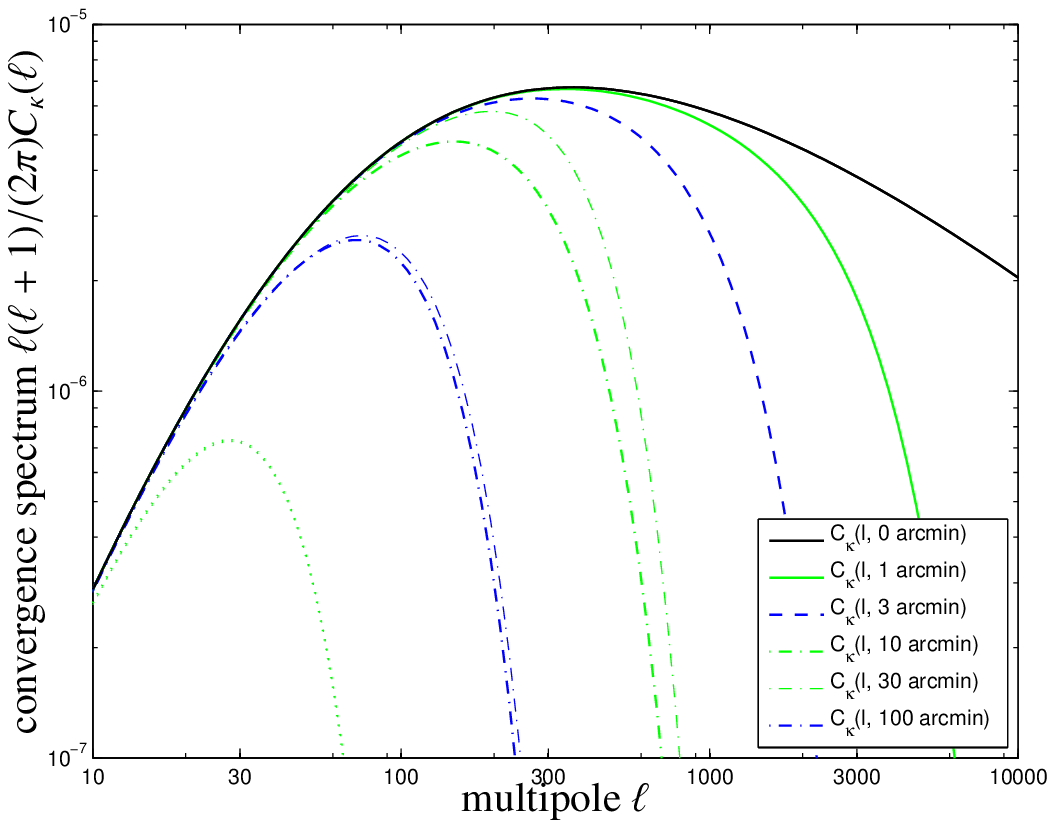}}
\caption{Angular convergence spectrum $C_\kappa(\ell)$ (black solid line) and with a Gaussian smoothing $W(\ell\theta)$ applied on a range of scales, $\theta=1,3,10,30,100$ arcmin. The faint lines for $\theta=10,30,100~\mathrm{arcmin}$ are derived with a nonlinear CDM spectrum, whereas the thick lines are computed with a linear CDM-spectrum.}
\label{fig_spectrum}
\end{center}
\end{figure}

For completeness we show the angular spectrum $C_\kappa(\ell)$ of the weak lensing convergence in Fig.~\ref{fig_spectrum} with a Gaussian smoothing $W(\ell\theta)$ applied on a range of scales $\theta$ which cuts off contributions on smaller multipoles $\ell$ with increasing $\theta$. From the smoothed spectrum we compute the smoothed convergence correlation functions $C_\kappa(\beta)$ by Fourier transform, and estimate in this way the correlation length of the convergence field $\kappa$. Furthermore, it gives the largest multipole $\ell$ for the numerical computation of the cumulants $\kappa_3$ and $\kappa_4$ needed at a given smoothing scale. Differences between spectra computed for linear and nonlinear CDM-spectra are small if $\theta$ is chosen large enough.

\bsp

\label{lastpage}

\end{document}